\documentclass{llncs}

\usepackage[]{algorithm2e}
\usepackage{algorithmic}
\usepackage{algorithm2e}
\usepackage{amsmath}
\usepackage{graphicx}
\usepackage{times}                               
\usepackage{graphicx}
\usepackage{tabularx}
\usepackage{mathtools}
\usepackage{array}
\usepackage{amsmath}
\usepackage{amssymb}
\usepackage{units}
\usepackage{textcomp,xspace}
\usepackage{enumerate}
\usepackage{multirow}
\usepackage{multicol}
\usepackage{subfigure}
\usepackage{color}

\begin{document}

\title{Achieving Efficient Realization of Kalman Filter on CGRA through Algorithm-Architecture Co-design}

\author{Farhad Merchant\inst{1} \and Tarun Vatwani\inst{2}
Anupam Chattopadhyay\inst{2} \and Soumyendu Raha\inst{3} \and S K Nandy\inst{4} \and Ranjani Narayan\inst{5}}

\institute{Institute for Communication Technologies and Embedded Systems, RWTH Aachen University, Germany\\
\and
Hardware and Embedded Systems Lab, Nanyang Technological University, Singapore
\and Scientific Computation Laboratory, Indian Institute of Science, Bangalore
\and Computer Aided Design Laboratory, Indian Institute of Science, Bangalore
\and Morphing Machines Pvt. Ltd. 
}

\maketitle

\begin{abstract}
In this paper, we present efficient realization of Kalman Filter (KF) that can achieve up to 65\% of the theoretical peak performance of underlying architecture platform. KF is realized using Modified Faddeeva Algorithm (MFA) as a basic building block due to its versatility and REDEFINE Coarse Grained Reconfigurable Architecture (CGRA) is used as a platform for experiments since REDEFINE is capable of supporting realization of a set algorithmic compute structures at run-time on a Reconfigurable Data-path (RDP). We perform several hardware and software based optimizations in the realization of KF to achieve 116\% improvement in terms of Gflops over the first realization of KF. Overall, with the presented approach for KF, 4-105x performance improvement in terms of Gflops/watt over several academically and commercially available realizations of KF is attained. In REDEFINE, we show that our implementation is scalable and the performance attained is commensurate with the underlying hardware resources\footnote{The paper has been accepted in ARC 2018}. 

\end{abstract}

\section{Introduction}
Coarse Grained Reconfigurable Architectures (CGRAs) have been active topic of research due to their power performance and flexibility \cite{Merc1}. CGRAs are capable of domain customization and they are targeted to achieve performance of Application Specific Integrated Circuits (ASICs) and flexibility of Field Programmable Gate Arrays (FPGAs) through presence of ASIC-like structures \cite{Merc2}\cite{exp1}. Typically, CGRAs occupy middle ground between ASICs and FPGAs \cite{cgr2}\cite{cgr3}. Furthermore, CGRAs are preferred as a platform in the application domains like signal processing and automotive that are embedded in nature \cite{Merc1}. Acceleration of scientific code on CGRA with high precision floating point arithmetic is yet to be fully explored. There exist very few attempts in the literature where computations like double precision General Matrix Multiplication (dgemm), QR factorization (dgeqrf), and LU factorization (dgetrf) are accelerated on CGRAs. Scientific application like Kalman Filter (KF) is rich in these matrix operations and has wide range of applications from trajectory optimization, and navigation to econometric \cite{alok1}. Here we see an opportunity to accelerate KF using a highly efficient Dense Linear Algebra (DLA) accelerator. Since, KF can be transformed as a series of matrix operations, we use Modified Faddeeva Algorithm (MFA) as a basic building block for our realization of KF \cite{faddeev1}. We take an approach of algorithm-architecture co-design where we identify several macro operations in the routines required for MFA and realize them on a specialized data path that leads to significant performance improvement in KF. Major Contributions in this paper are as follows:

\begin{itemize}
\item We adopt a library based approach and realize KF using MFA where MFA is realized using dgemm, dgetrf, and dgeqrf routines. The first realization is capable of achieving 30\% of the theoretical peak performance of the underlying platform. We call this implementation as a base realization of KF
\item A set of macro operations identified in dgemm, dgeqrf, and dgetrf are realized on a tightly coupled Reconfigurable Data-path (RDP) and the implementation results in 66\% of performance improvement over base realization at minimal area and energy cost. We call this implementation as hardware optimized KF
\item A scheduling optimization is presented for KF that results in 30\% performance improvement over the hardware optimized realization and 116\% improvement over base realization. We also show that the final implementation is able to achieve 4-105x performance improvement over several academic and commercial realizations of KF. We call this implementation as software optimized KF 
\item Parallel realization of KF is presented on REDEFINE and it is shown that REDEFINE scales well with increasing size of co-variance matrix
\end{itemize}

For our experiments, we use Processing Element (PE) design presented in \cite{Merc1}. We optimize Floating Point Unit (FPU) design presented in \cite{fpu2} with recommendations presented in \cite{fpu3} for optimum Instructions Per Cycle (IPC). 
The paper is organized as follows: In section \ref{sec:rw}, MFA, KF and REDEFINE are discussed. In section \ref{sec:mot}, we present case studies on multicore and General Purpose Graphics Processing Unit (GPGPU) and show that the performance attained on these platforms even with highly tuned software packages is not satisfactory. KF implementation is discussed in section \ref{sec:kf}. Parallel realization of KF and results are discussed in section \ref{sec:res}. We conclude our work in section \ref{sec:con}.

\section{Background and Related Work}\label{sec:rw}
In this section, we first discuss MFA and KF briefly and show connection between MFA and KF. In section \ref{sec:redefine}, we introduce REDEFINE and discuss unique features of REDEFINE CGRA. We focus on some of the recent realizations of KF in section \ref{sec:related_work}. 
\subsection{Background}
\subsubsection{Modified Faddeeva Algorithm and Kalman Filter}\label{sec:back_kal}
MFA was originally presented in \cite{faddeev1} that is an enhancement over the original Faddeeva algorithm. QR factorization was incorporated instead of Gaussian elimination process to improve stability of Faddeeva algorithm. For rectangular matrices $A_{m\times n}, B_{m\times p}, C_{k\times n}$, and $D_{k\times p}$, compound matrix is shown in equation \ref{eqn:eqn1}.
\begin{align}\label{eqn:eqn1}
M = \begin{bmatrix} A & B \\ -C & D\end{bmatrix}
\end{align}
Applying MFA on compound matrix $M$ incorporates following two steps. 
\begin{itemize}
	\item {\bf Step 1:} Upper triangularization of matrix $A$ using QR factorization and update matrix $B$
		\item {\bf Step 2:} Annihilate matrix $C$ using diagonal elements of upper traingularized matrix $A$
\end{itemize}

\begin{figure*}[!ht]
\centerline{\subfigure[Several Operations that can be Performed Using MFA based on Different Initial Values of Matrices $A,B,C$, and $D$]{\includegraphics[scale=0.17]{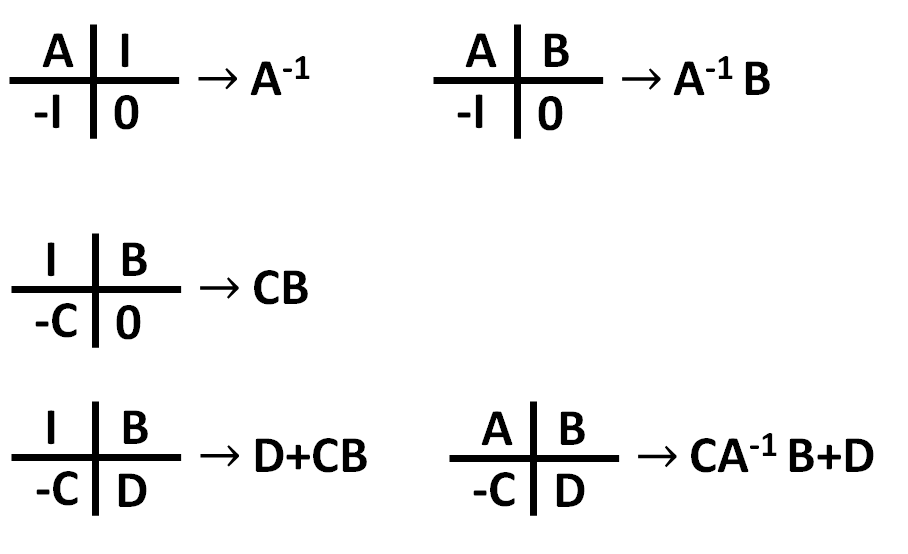}%
\label{fig:mfa1}}
\subfigure[Multi Dimensional Kalman Filter]{\includegraphics[scale=0.17]{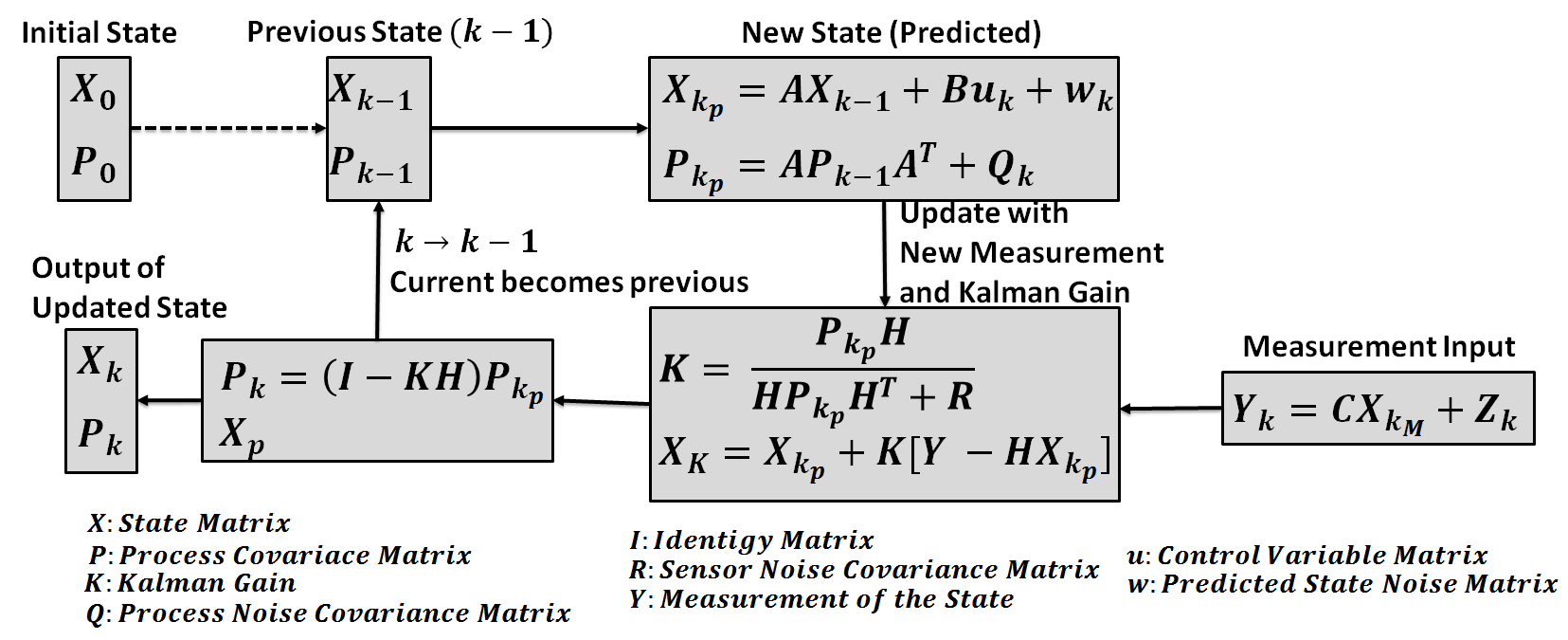}%
\label{fig:kalman1}}} 
\caption{MFA and KF}
\end{figure*}
%

At the end of an MFA operation, the matrix $D$ is of the form $D+CA^{-1}B$ that is also known as the Schur complement. Versatility of MFA is described in figure \ref{fig:mfa1} where it is shown that based on different initial input matrices, several operations like matrix multiplication, matrix addition, and linear system solution can be obtained. Due to versatility in MFA, it is desirable to write applications in terms of matrix operation that can translate into series of calls of MFA. One such application is KF. A simplistic multi-dimensional KF is shown in the figure \ref{fig:kalman1}. 


An iteration of KF requires complex matrix operations ranging from matrix multiplication to computation of numerical stable matrix inverse. One such classical GR based numerical stable approach is presented in \cite{Kal1}. From figure \ref{fig:kalman1}, matrix inverse being the most complex operation in KF, and computation of inverse using QR factorization being a numerical stable process, proposed library based approach is the most suitable for such applications. 

\subsubsection{REDEFINE}\label{sec:redefine}
\begin{figure*}[!ht]
	\begin{centering}
	\includegraphics[scale=0.20]{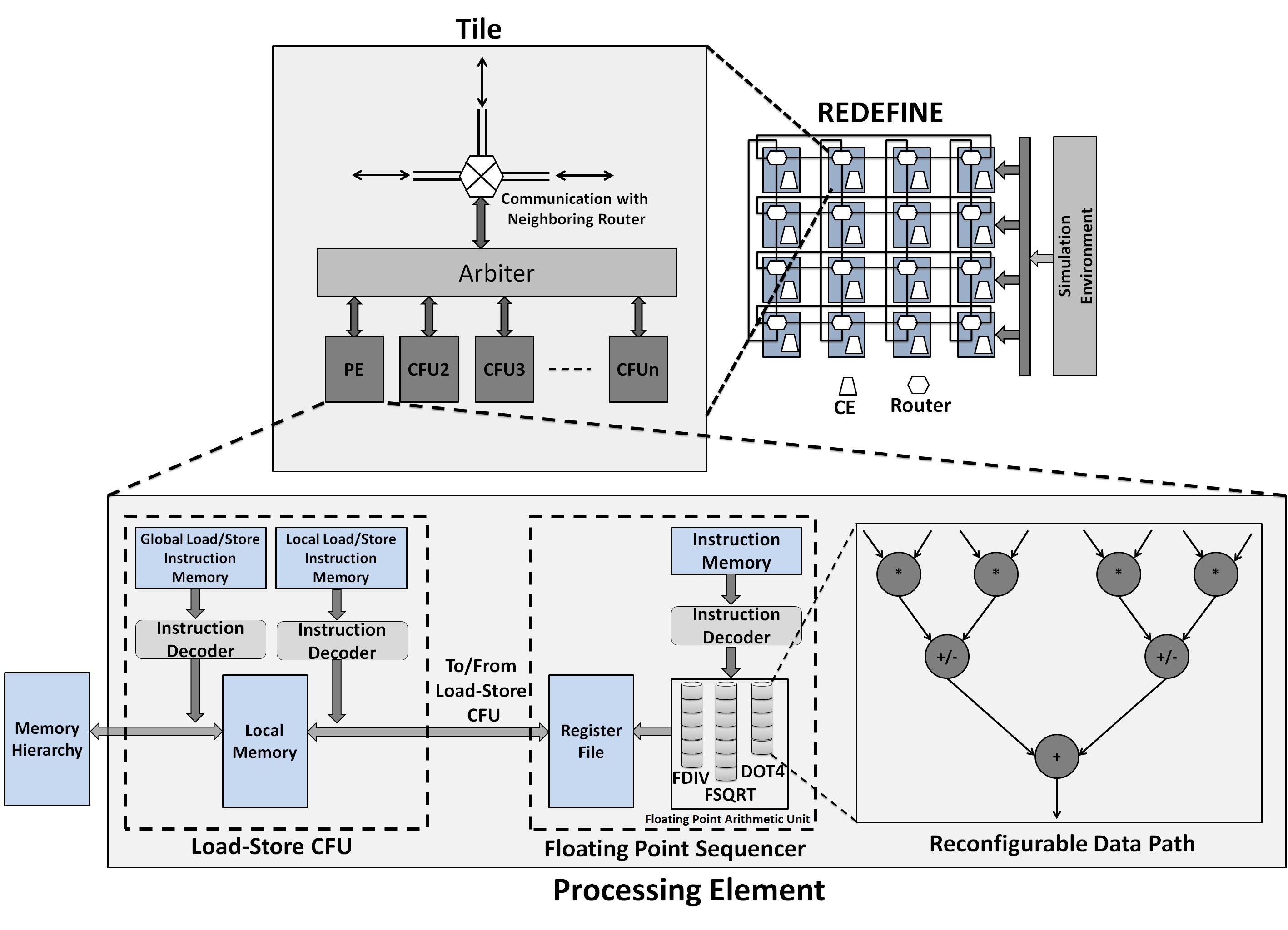}
	\caption{REDEFINE CGRA and Design of PE (with RDP)}
	\label{fig:pe1}
	\end{centering}
\end{figure*}
REDEFINE is a CGRA that can be domain customized for variety application domains \cite{tpds1}.In REDEFINE, several Tiles are connected through a packet switched Network-on-Chip (NoC) \cite{cgr1}\cite{tpds1}. Each Tile consists of a Compute Element (CE) and a Router. CEs in REDEFINE can be enhanced with a Custom Function Units (CFUs) that can be tailored for a desired application domain. For our implementation, we use a Processing Element (PE) that is highly customized for Dense Linear Algebra (DLA) computations \cite{Merc1}. REDEFINE micro-architecture along with PE is shown in figure \ref{fig:pe1}. It can be observed in the figure \ref{fig:pe1} that the PE consists of a RDP as an arithmetic unit that can be reconfigured at run-time through instruction to realize identified macro operations in the Directed Acyclic Graph (DAG) of DLA computations. It is well established in the literature that such an approach yields significant improvement in energy efficiency in the overall system \cite{hyper1}.

\subsection{Related Work}\label{sec:related_work}
Due to wide range of applications of KF, there are several implementations of KF in the literature. We present brief summary of some of the recent realizations of KF. Early realizations of KF are based on systolic array since systolic arrays are highly suitable for matrix computations \cite{Kung2}. One of the recent realization of KF presented in \cite{kalman_app4} is targeted for advanced driver assistance system. The implementation presented in \cite{kalman_app4} does not focus on the acceleration of matrix operations encountered in KF and hence leaves opportunities for further performance improvements. Similarly, KF implementation presented in \cite{kal_new1} focuses on parallel implementation using OpenMP and FPGA based implementation presented in \cite{kal_new2} focuses on optimizations in arithmetic operations in KF for performance improvement. MFA based KF presented in \cite{kal_latest1} focuses on denoising of images with additive white Gaussian noise. Although the work presented in \cite{kal_latest1} realized KF using MFA, the implementation is not scalable and can operate only on the images of size $512\times 512$. Furthermore, focus of the work presented in \cite{kal_latest1} is not on acceleration of KF through acceleration of matrix operations in MFA. In this paper, we focus on energy efficient fast realization of KF by accelerating gemm, QR factorization, and LU factorization.   

\section{Case Studies}\label{sec:mot}
\begin{figure*}[!ht]
\centerline{\subfigure[Percentage of Theoretical Peak Performance Attained in dgemm, dgeqrf, and dgetrf in Intel Core i7 (PLASMA) and Nvidia Tesla C2075 (MAGMA)]{\includegraphics[scale=0.17]{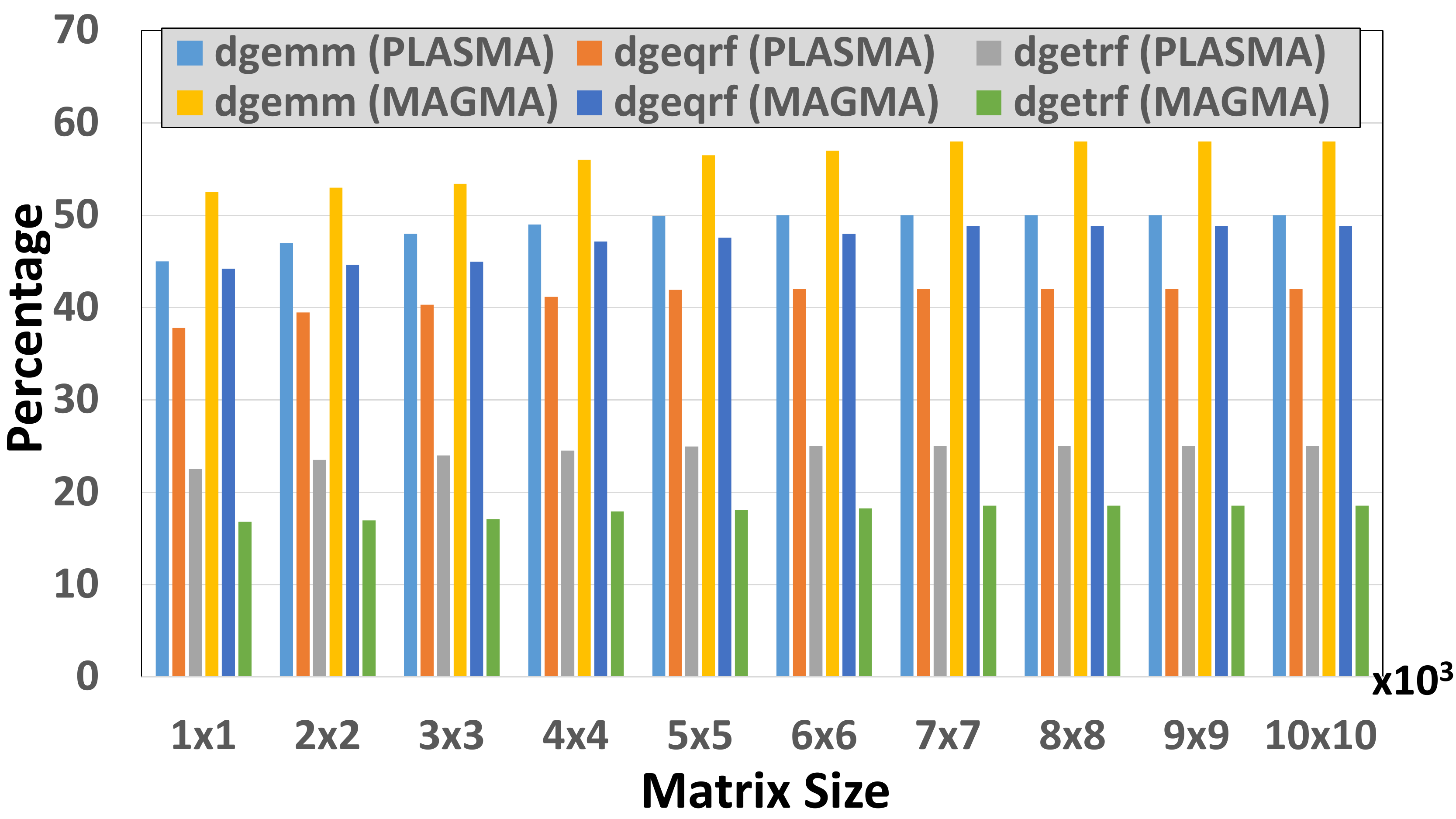}%
\label{fig:case1}}
\subfigure[Performance Attained in terms of Gflops/watt in dgemm, dgeqrf, and dgetrf in Intel Core i7 (PLASMA) and Nvidia Tesla C2075 (MAGMA)]{\includegraphics[scale=0.17]{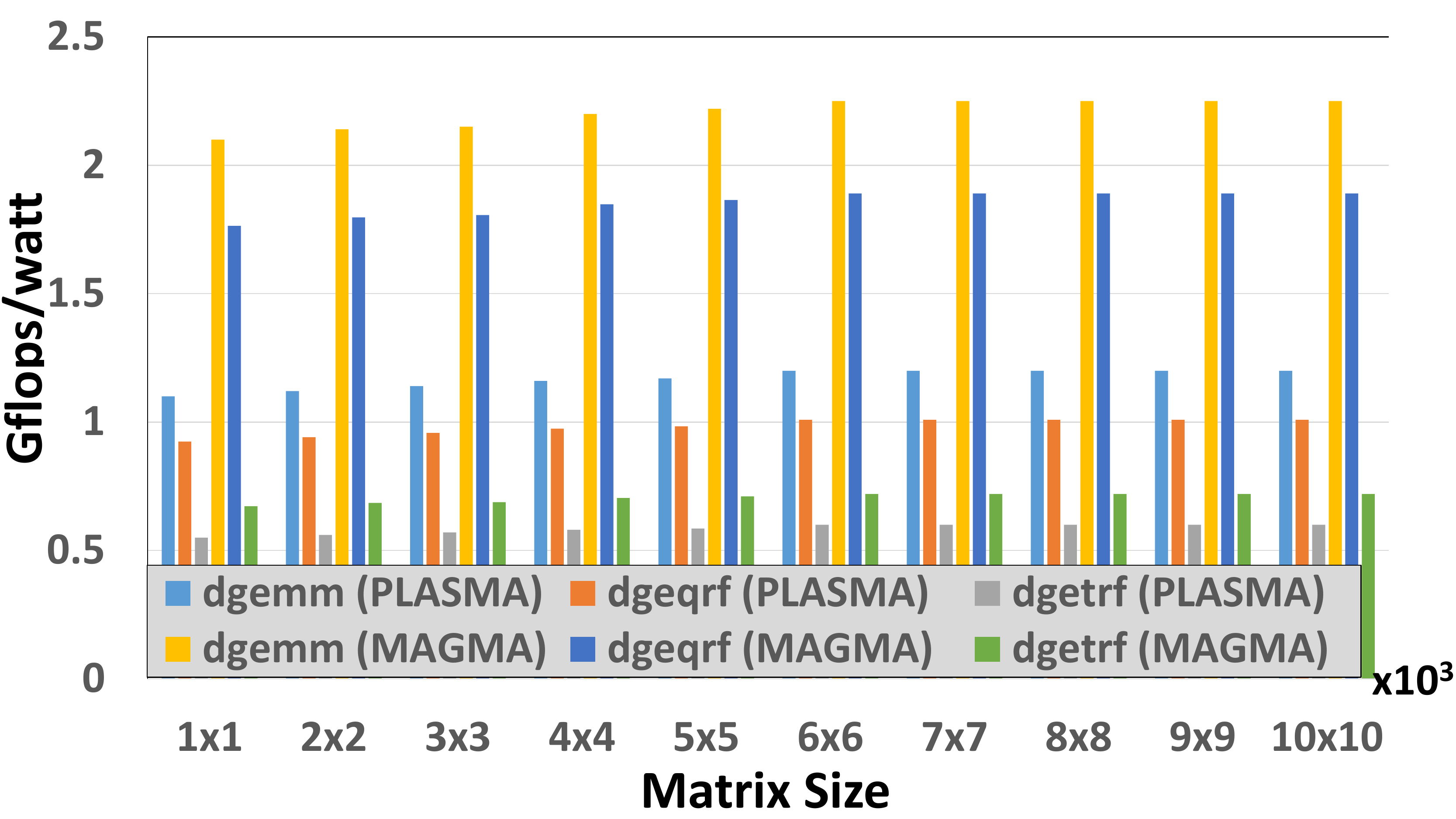}%
\label{fig:case2}}} 
\caption{Performance of dgemm, dgeqrf, dgetrf, and KF in multicore and GPGPU}
\label{fig: res3}
\end{figure*}
\begin{figure*}
\centerline{\subfigure[Percentage of Theoretical Peak Performance Attained in KF in Intel Core i7 (PLASMA) and Nvidia Tesla C2075 (MAGMA) ]{\includegraphics[scale=0.17]{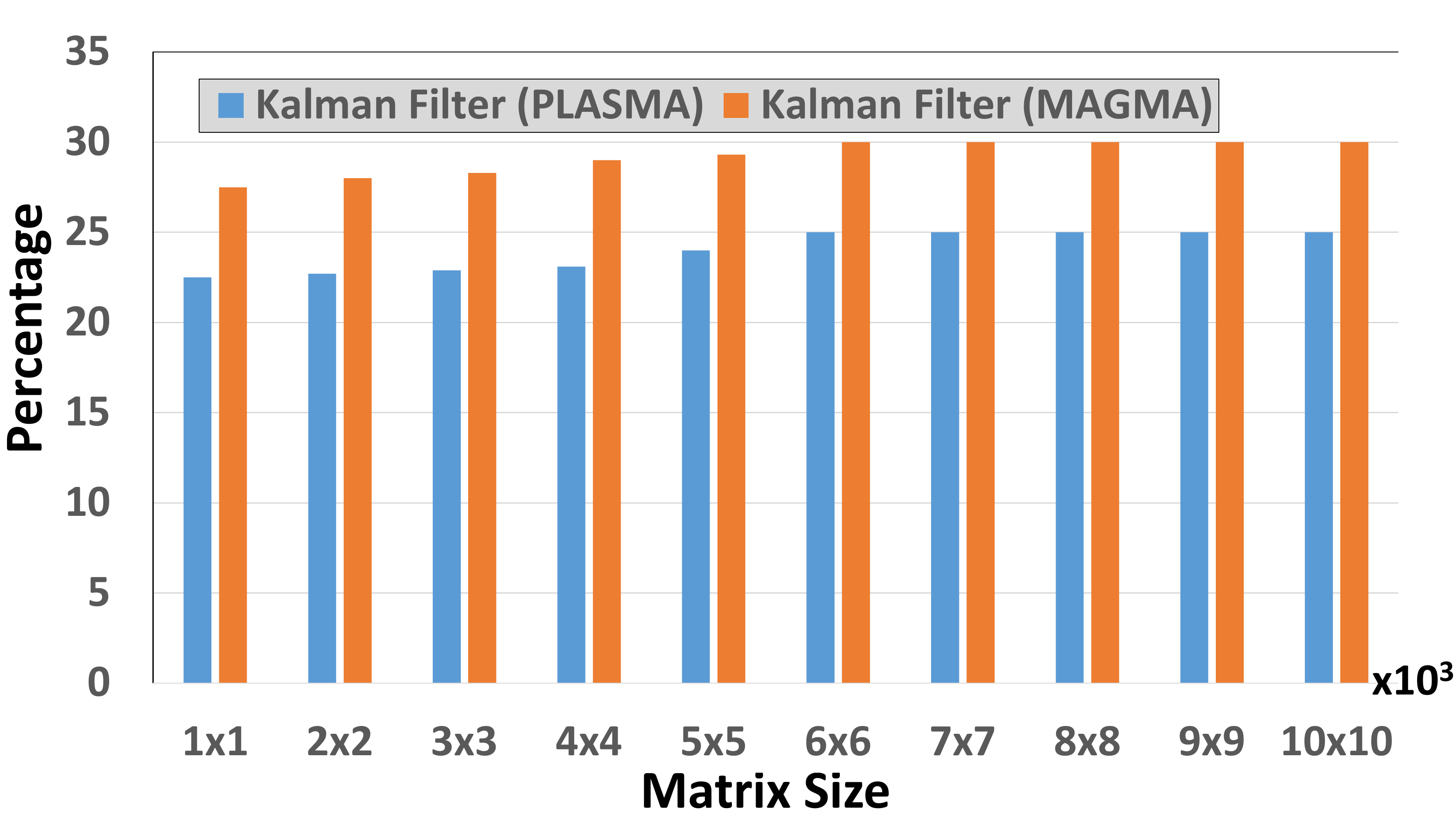}%
\label{fig:case3}}
\subfigure[Performance Attained in terms of Gflops/watt in KF in Intel Core i7 (PLASMA) and Nvidia Tesla C2075 (MAGMA)]{\includegraphics[scale=0.17]{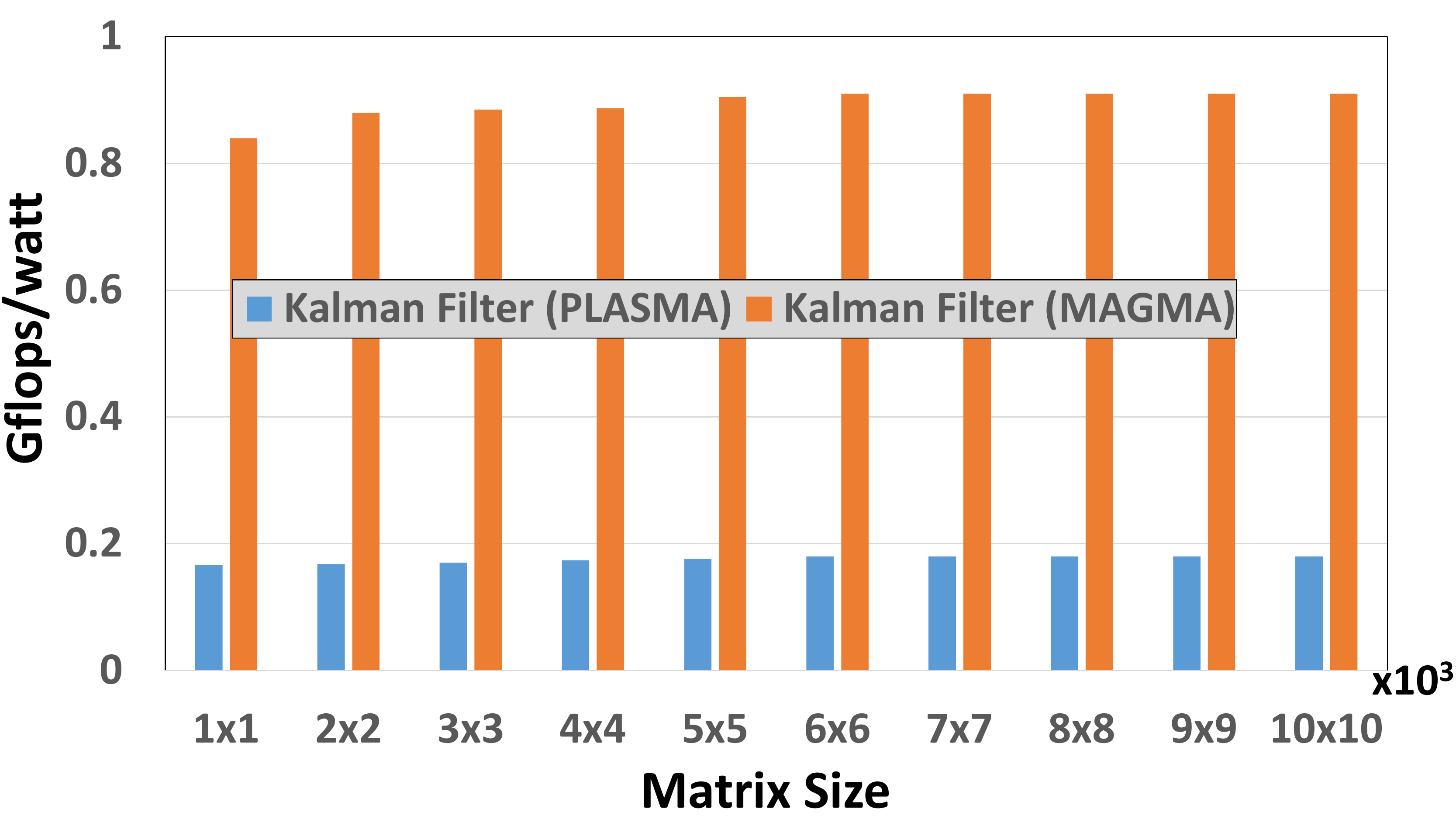}%
\label{fig:case4}}}
\caption{Performance of dgemm, dgeqrf, dgetrf, and KF in multicore and GPGPU}
\label{fig: res3}
\end{figure*}
In this section we present case studies on the operation encountered in MFA. Since, we implement KF using double precision FPU, we present case studies on dgeqrf, dgetrf, and dgemm. We discuss performance of dgemm, dgeqrf, and dgetrf in highly optimized software libraries like Parallel Linear Algebra Software for Multicore Architectures (PLASMA) and Matrix Algebra for GPU and Multicore Architectures (MAGMA).


\subsection{dgeqrf}
Routine dgeqrf computes QR factorization of an input matrix $A$ and returns uppter triangular matrix $R$ and $Q$ where $QQ^T = Q^TQ = I$. Householder Transform (HT) based QR factorization routine in Linear Algebra Package (LAPACK) is shown in algorithm \ref{algo:dgeqr2}.
\begin{algorithm}
\KwIn{Matrix $A_{m\times n}$}
\KwOut{Upper triangle matrix $R$}
\For{$i=1$ to $n$}{
       Compute Householder vector v\\
   	Compute $P$ where $P = I - 2vv^T$\\
	Update trailing matrix  using $dgemv$
}
\caption{dgeqr2 in LAPACK (dgemv is double precision matrix-vector multiplication)}
\label{algo:dgeqr2}
\end{algorithm}
\begin{algorithm}
\KwIn{Matrix $A_{m\times n}$}
\KwOut{Upper triangle matrix $R$}
\For{$i=1$ to $n$}{
       Compute Householder vectors for block column $m\times k$\\
   	Compute $P$ where $P$ is multiplication of Householder vectors \\
	Update trailing matrix  using $dgemm$
}
\caption{dgeqrf in LAPACK}
\label{algo:dgeqrf}
\end{algorithm}

In algorithm \ref{algo:dgeqr2}, majority of the computations are performed in terms of double precision matrix vector multiplication (dgemv). Since dgemv is a memory bound operation, the realization of dgeqr2 yields hardly 10\% of the theoretical peak performance in GPGPU and 2-3\% of the theoretical peak performance in multicores. 
In the dense linear algebra software packages like LAPACK, PLASMA, and MGAMA, the software routines are tiled/blocked and written in terms of dgemm for efficient exploitation of memory hierarchy of the underlying platform where dgemm is compute bound operation \cite{magma1}. dgeqrf routine that uses dgeqr2 and dgemm as subroutines is shown in algorithm \ref{algo:dgeqrf}.

\subsection{dgetrf}
dgetrf routine computes LU factorization of a general $m\times n$ matrix, so $A = PLU$ where $L$ is unit lower triangular, $U$ is upper triangular matrix, and $P$ is permutation matrix. dgetrf2 routine is shown in algorithm \ref{algo:dgetrf2}. 

\begin{algorithm}
\KwIn{Matrix $A_{m\times n}$}
\KwOut{Unit lower triangular matrix $L$ and upper triangular matrix $U$}
\For{$i=1$ to $n$}{
    Find pivot \\
    Interchange pivot rows \\
    Compute elements of $L$ matrix \\
    Update the trailing matrix using dgemv
}
\caption{dgetrf2 in LAPACK}
\label{algo:dgetrf2}
\end{algorithm}

\begin{algorithm}
\KwIn{Matrix $A_{m\times n}$}
\KwOut{Unit lower triangular matrix $L$ and upper triangular matrix $U$}
\For{$i=1$ to $n$}{
    Find pivot \\
    Interchange pivot rows \\
    Compute elements of $L$ matrix \\
    Update the trailing matrix using dgemm
}
\caption{dgetrf in LAPACK}
\label{algo:dgetrf}
\end{algorithm}

A similar approach to QR factorization is adopted for LU factorization in LAPACK where a blocked routine shown in the algorithm \ref{algo:dgetrf} is developed to exploit the memory hierarchy of cache based platforms. For our implementation, since we do not require pivoting, we remove pivoting related routine from dgetrf2 and adopt the routine to support realization of MFA.

\subsection{dgemm}
 
Due to prevalent in the literature, we do not reproduce pseudo code of dgemm. Typically, dgemm is part of Level-3 Basic Linear Algebra Subprograms (BLAS) and used as a subroutine in the operations like LU and QR factorizations since dgemm is compute bound operation and theoretically dgemm is capable of attaining $O(n)$ computations to communication ratio for the matrices of size $n\times n$ \cite{nick1}.   

\subsection{Performance Evaluation of dgeqrf, dgetrf and dgemm on multicore and GPGPU}

We evaluate performance of dgeqrf, dgetrf, and dgemm on multicore and GPGPU. Performance in terms of percentage of theoretical peak performance in Intel Core i7 for dgeqrf, dgetrf, and dgemm is shown in figure \ref{fig:case1}. It can be observed in the figure \ref{fig:case1} that in Intel Core i7, performance attained by dgeqrf is 42\% (21 Gflops), performance attained by dgetrf is 25\% (12 Gflops) and performance attained by dgemm is 50\% (24 Gflops). Similarly, percentage of theoretical peak performance attained in GPGPU for dgeqrf is 48\% (247.2 Gflops), for dgemm it is 58\% (298 Gflops), and for dgetrf it is 20\% (103 Gflops). We ensure to compile PLASMA and MAGMA with OpenBLAS \cite{augem1}. Performance attained in terms of Gflops/watt ranges between 0.6 to 2.1 for GPGPU and 0.6 to 1.1 for multicore as shown in figure \ref{fig:case2}.  It can be observed that the performance attained by dgeqrf is 84.2\% of the performance attained by dgemm and performance attained by dgetrf is 50\% of the performance attained by dgemm. This is mainly due to presence of non-parallelizable computations in dgeqrf and dgetrf that limits the performance considering Amdahl's law \cite{Amdahl1}. 

Similarly, for KF the performance attained in terms of percentage of theoretical peak performance is 25\% for multicore and 30\% for GPGPU as shown in figure \ref{fig:case3}. Performance attained in terms of Gflops/watt for KF is 0.15 in multicore and 0.9 in GPGPU.  
Based on our case studies in dgemm, dgeqrf, and dgetrf it can be inferred that the performance attained in the latest multicore and GPGPU is hardly 50-52\% even for highly parallel operations like dgemm. Due to this shortcoming of multicore and GPGPU, we choose a customizable platform for our implementation presented in \cite{Merc1} that is capable of achieving up to 74\% of the theoretical peak in dgemm \cite{Merc1}\cite{Merc2}.

\section{Kalman Filter Realization in Processing Element}\label{sec:kf}
We present realization of KF on PE depicted in the figure \ref{fig:pe1}. Three different implementations of KF are presented here: 1) base implementation of KF, 2) hardware optimized KF, and 3) software optimized KF.

\subsection{Base Implementation of KF}
In base implementation of KF, we realize matrix operation of in the MFA using scalar multiplier and adder in the PE. In base realization of KF, we are able to achieve up to 30\% of the theoretical peak performance of the PE. Here, since we are using only one multiplier and adder for our implementation, the theoretical peak performance of the PE is 1.4 Gflops at 700 MHz.  

\subsection{Hardware Optimized KF}


\begin{figure*}[!ht]
\centerline{\subfigure[Different Configurations of RDP Corresponding to Identified Macro Operations in dgemm, dgeqrf, and dgetrf]{\includegraphics[scale=0.16]{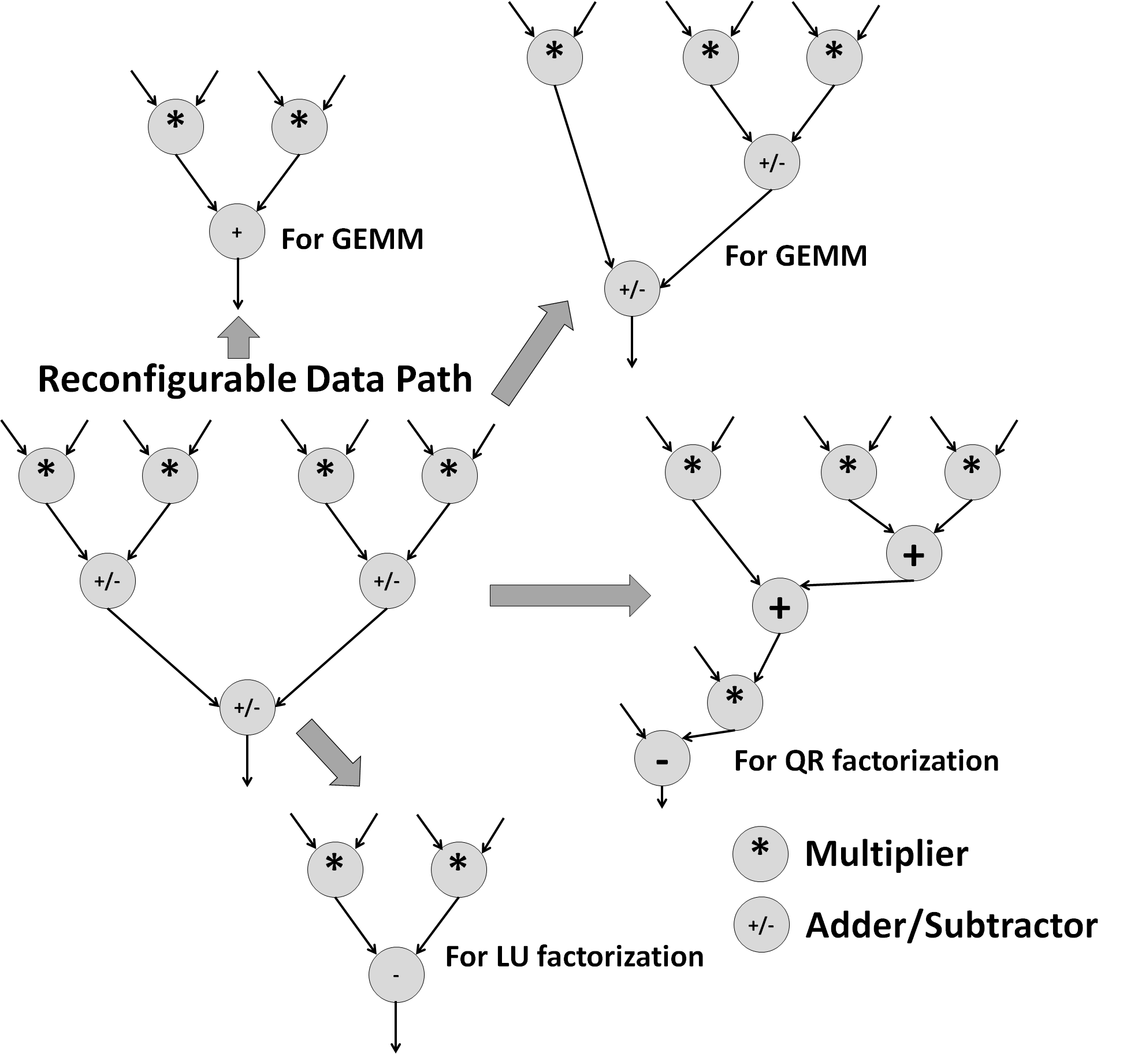}%
\label{fig:config1}}
\subfigure[Software Optimization in KF Resulting in Reduction in the Run-time (logical diagram)]{\includegraphics[scale=0.16]{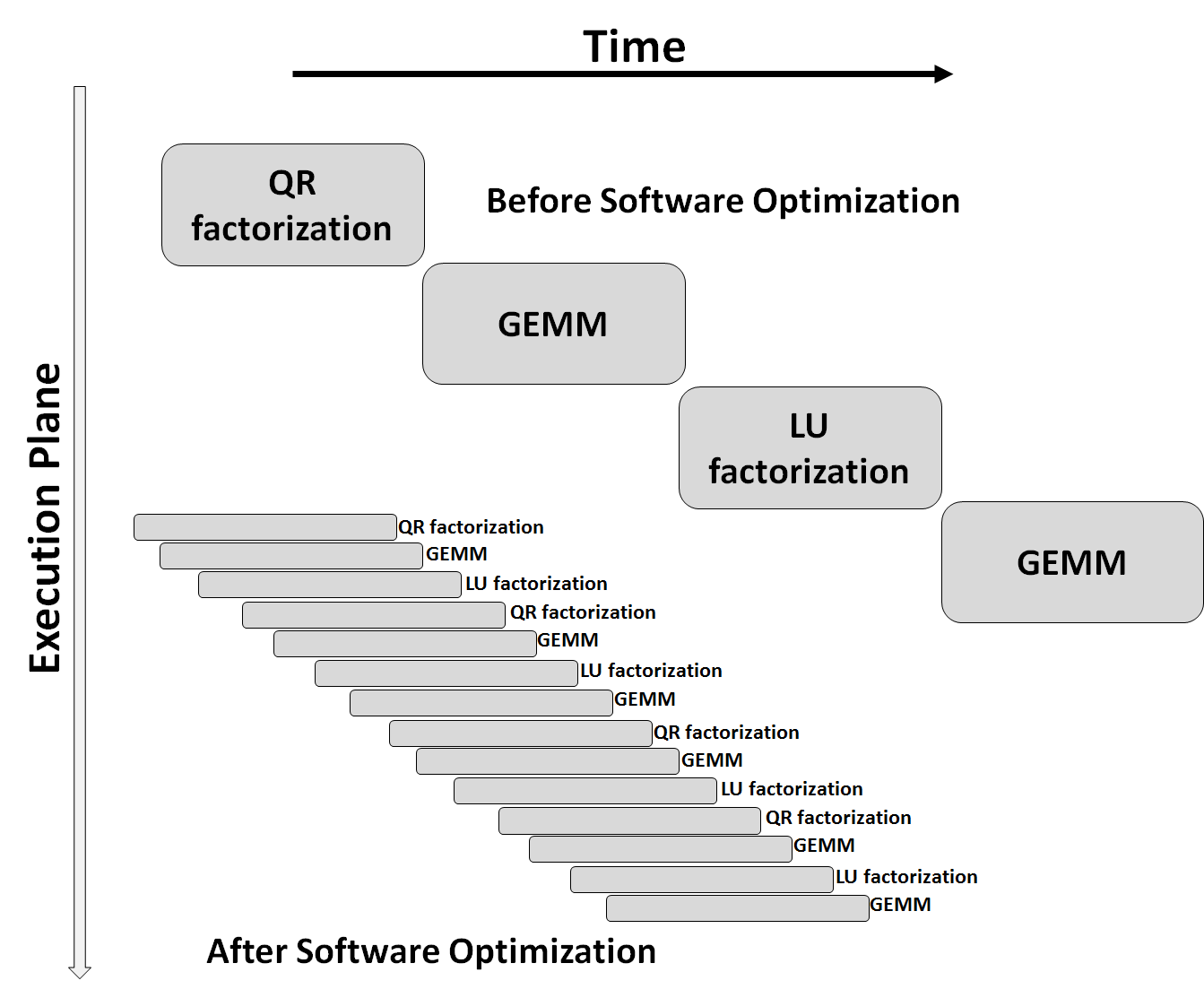}%
\label{fig:soft_opt}}} 
\caption{RDP Configurations for KF and Scheduling of Different Routines in KF}
\end{figure*}

In hardware optimized KF, we revisit the basic operations required to be performed in MFA like dgemm, dgeqrf, and dgetrf and identify several macro operations in these basic operations. We realize these macro operations in RDP depicted in the figure \ref{fig:pe1}. Configurations of RDP corresponding to the identified macro operations in dgemm, dgeqrf and dgetrf are shown in figure \ref{fig:config1}. With these configurations, we achieve up to 50\% of the theoretical peak of the PE where theoretical peak of the PE is 4.9 Gflops at 700 MHz. Here theoretical peak of the PE is increased since we are using RDP that consists of 4 multipliers and 3 adders.   

\begin{figure}[!ht]
	\begin{centering}
	\includegraphics[scale=0.25]{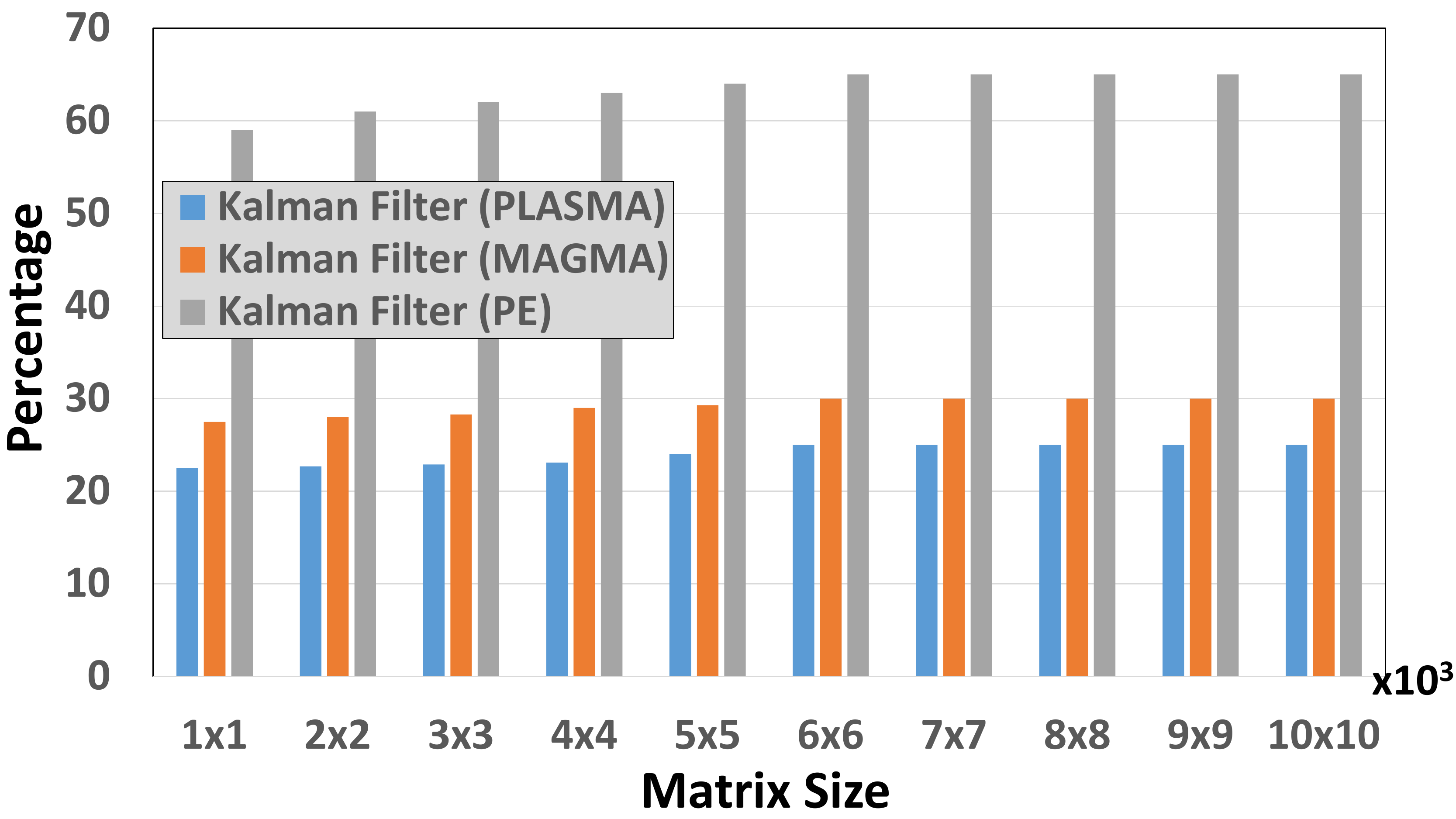}
	\caption{Performance of KF in PE, multicore, and GPGPU}
	\label{fig:perf1}
	\end{centering}
\end{figure}

We perform a software optimization in MFA by analysis of the DAG of MFA. We overlap dgeqrf, dgemm, and dgetrf routines as shown in figure \ref{fig:soft_opt}. Optimization diagram shown in the figure \ref{fig:soft_opt} is logical flow of computations post software optimization. To overlap these routines, we identify pipeline stalls in the PE while execution and insert independent instructions while also maintaining operation correctness. Overlapping dgeqrf, dgemm, and dgetrf results in significant reduction in the run-time of MFA that directly translates to performance improvement in KF. After software optimization, we are able to attain 65\% of the theoretical peak in PE which is 30\% improvement. Performance improvement after each optimization is shown in figure \ref{fig:perf1}. In the figure \ref{fig:perf1}, we have also incorporated the performance attained in multicore and GPGPU.   

It can be observed in the figure \ref{fig:perf1} that the performance of KF in multicore and GPGPU is hardly 20-30\% of the theoretical peak of these platforms, while we achieve up to 65\% of the theoretical peak of PE in KF which is 2.15x higher.

\section{Parallel Realization and Results}\label{sec:res}
\begin{figure}[!ht]
	\begin{centering}
	\includegraphics[scale=0.30]{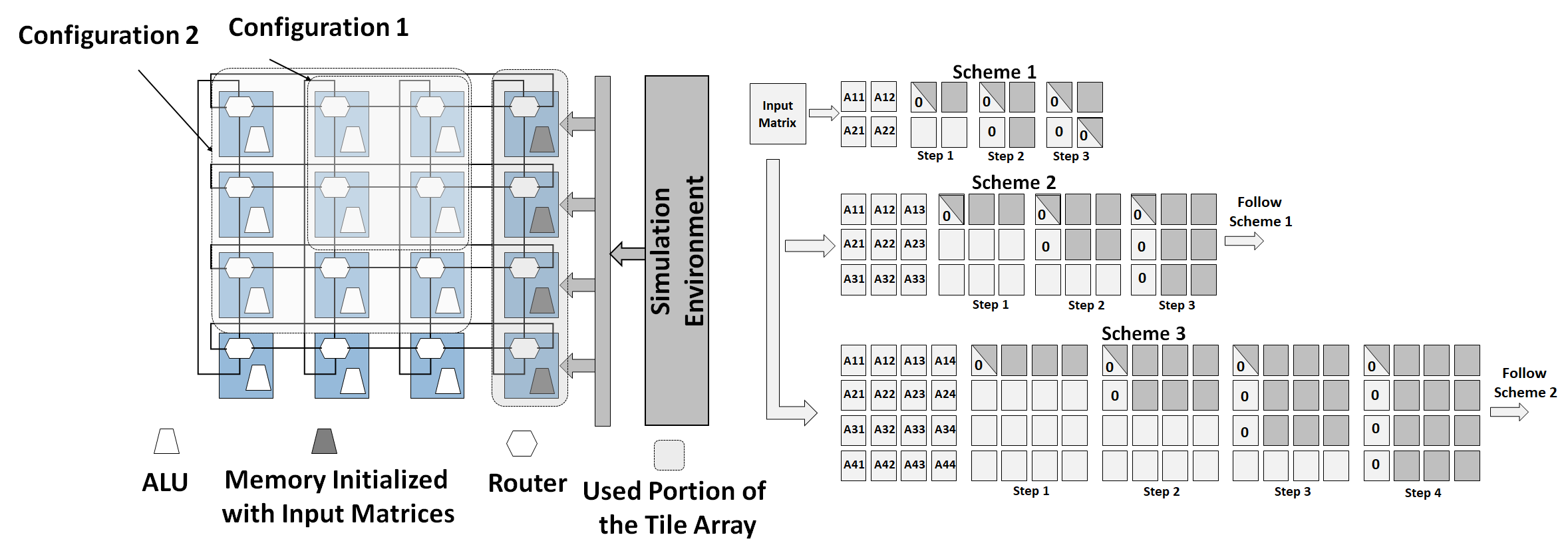}
	\caption{Different Configurations of REDEFINE for KF Realization and Scheduling}
	\label{fig:redefine_config}
	\end{centering}
\end{figure}

For parallel realization of KF, we use three different configurations of REDEFINE. Two configurations are shown in figure \ref{fig:redefine_config}. In configuration $1$ we use $2\times 2$ Tile array, in configuration $2$ we use $3\times 3$ Tile array, and in configuration $3$, we use $4\times 4$ Tile array. In our simulations, for configuration $1$ and configuration $2$, we use last column of the Tile array as a memory where we attach memories as a PE. 

\begin{figure*}[!ht]
\centerline{\subfigure[Performance in Different Configurations of REDEFINE in KF Realization]{\includegraphics[scale=0.17]{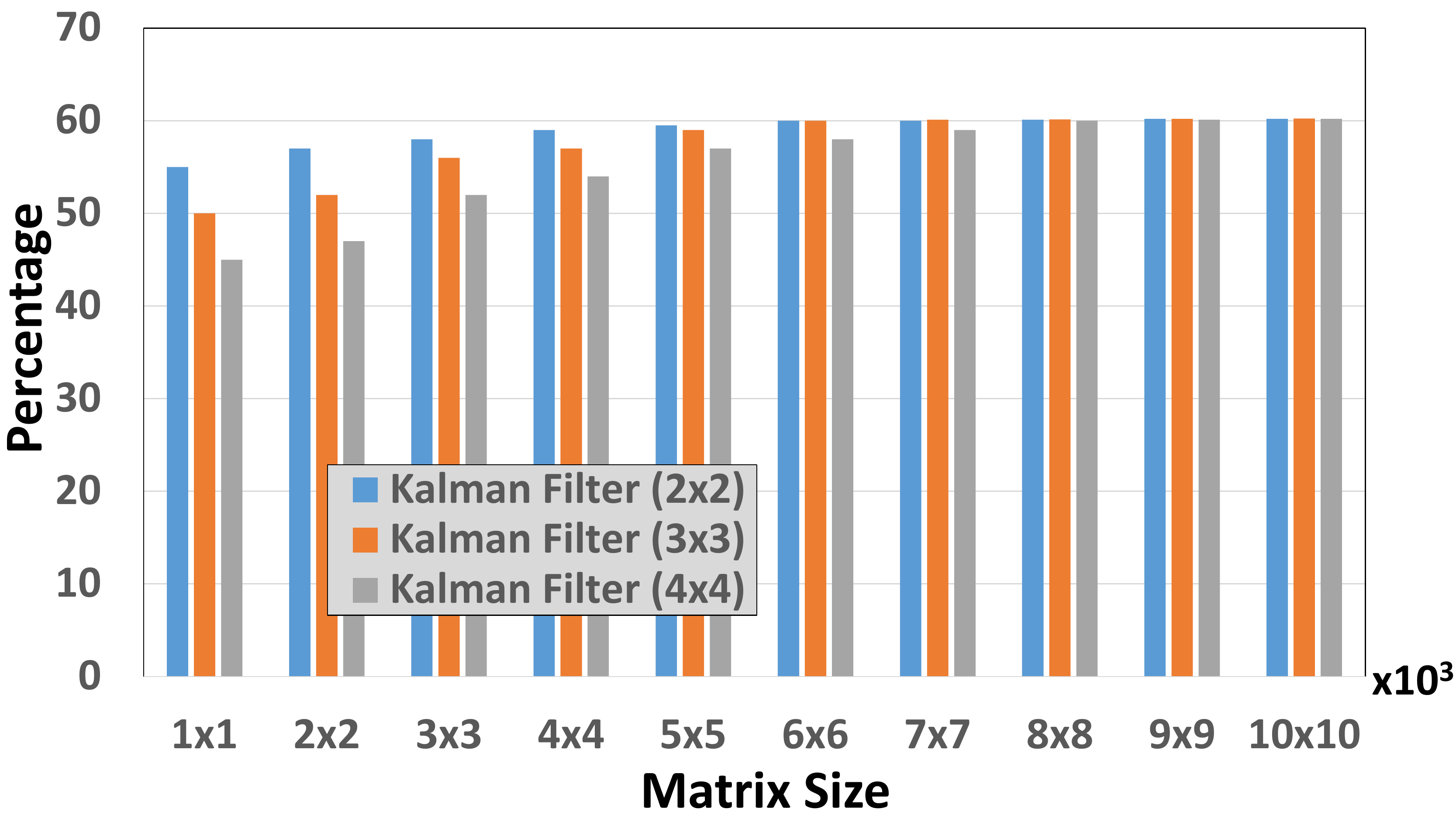}%
\label{fig:perf_redefine}}
\subfigure[Power Performance Comparison of PE with Other Platforms for KF]{\includegraphics[scale=0.17]{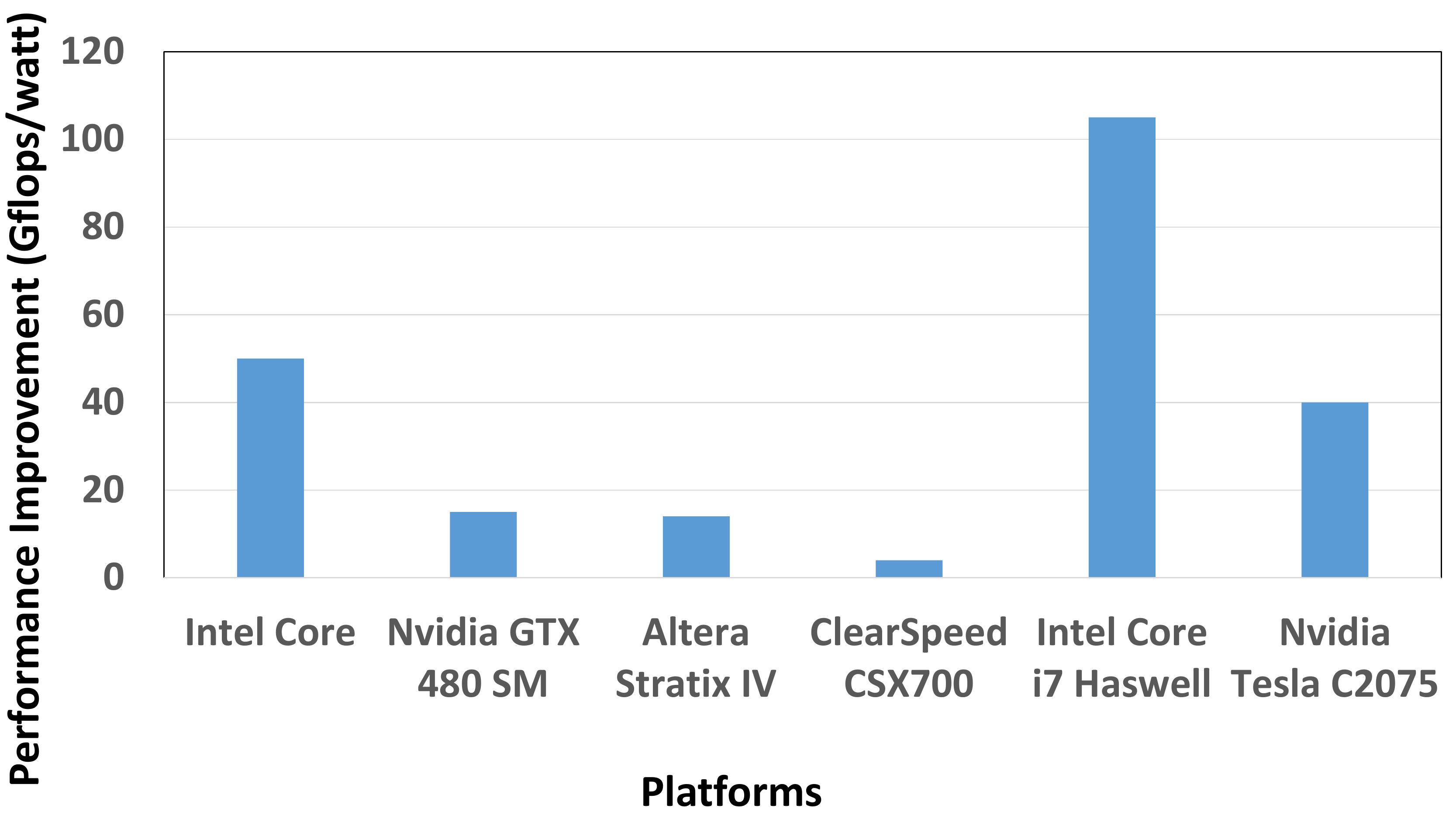}%
\label{fig:res1}}} 
\caption{RDP Configurations for KF and Scheduling of Different Routines in KF}
\end{figure*}


In configuration $3$, we use entire Tile array for computations and hence we attach another memory PE to the Router along side the compute PE. Memory PE is divided into to segments and the second half of the segment in memory PE acts as a global memory that is accessible by all other Tiles while the first half of the memory segment is private to the compute PE that is used for computations on local data. Typically, we have 256K bytes of memory per Tile and compute PE consists of 256 registers of width 64 bits. Scheduling technique for REDEFINE is shown in the right side of the figure \ref{fig:redefine_config}. We use a technique where we divide input matrix into $k\times k$ blocks where $k$ is size of row/column of the Tile array. These blocks are further divided into the sub-blocks where size of the sub-blocks depend on the number of local registers available and size of the local memory available to a PE. Blocks of the matrices are loaded and computation is performed and the result is stored to the global memory. Percentage of theoretical peak performance attained for each configuration is shown in figure \ref{fig:perf_redefine}. It can be observed in the figure \ref{fig:perf_redefine} that the performance attained for each configuration saturates at 60\% of the theoretical peak performance which is 2x higher than the performance attained in multicore and GPGPU. 


We evaluate power performance of PE based on technique presented in \cite{tpds1}. We compare PE with other platform for KF as shown in figure \ref{fig:res1}.	It can be observed that the PE is capable of achieving 4-105x higher performance improvement over platforms like ClearSpeed CSX700, Intel Core i7, and Nvidia GPGPU.

\section{Conclusion}\label{sec:con}
In this paper, we presented efficient realization of KF. We used versatile MFA as a tool for efficient realization of KF. Based on the case studies presented on dgemm, dgetrf, and dgeqrf, it was identified that the performance of these operations on multiore and GPGPU is not satisfactory even with hightly optimized software pachages like PLASMA and MAGMA. It was also shown that the performance attained in KF on multicore and GPGPU is 20-30\% of the theoretical peak performance of underlying platform. To accelerate KF on customizable platform like REDEFINE, we identify macro operations in the routines of MFA and realized them on RDP. Our approach resulted in 67\% improvement in the performance of KF. A software tuning in MFA was also presented that resulted in 30\% performance improvement over the hardware optimized KF. Overall, our approach of algorithm-architecture co-design resulted in 116\% of performance improvement over the base realization of KF. In terms of Gflops/watt, KF is 2-105x better than multicore and GPGPU platforms.   


\bibliographystyle{plain}
\bibliography{ref}

\end{document}